\documentclass[aps,twocolumn,showpacs,preprintnumbers,nofootinbib,prl,superscriptaddress,groupedaddress,10pt]{revtex4-1}

\makeatletter
\def\l@subsubsection#1#2{}
\def\l@subsubsubsection#1#2{}
\makeatother

\usepackage{graphicx,amssymb,amsmath,amsthm,amsfonts,epsfig,epsf}
\usepackage[usenames]{color}
\usepackage{epstopdf}
\definecolor{darkred}{rgb}{0.5,0,0}
\usepackage{bm}
\usepackage{dcolumn}
\usepackage[latin1]{inputenc}
\usepackage{latexsym}
\usepackage{rotating}
\usepackage{longtable}

\usepackage{enumerate}
\usepackage{tensor,multirow}
\usepackage{url,pifont}

\usepackage{gensymb}

\usepackage[linktocpage]{hyperref}

\def\be{\begin{equation}}
\def\ee{\end{equation}}
\newcommand{\beq}{\begin{eqnarray}}
\newcommand{\eeq}{\end{eqnarray}}

\def\ba{\begin{align}}
\def\ea{\end{align}}

\begin{document}



\title{Tests for the existence of horizons through gravitational wave echoes}

\author{Vitor Cardoso$^{1,2}$, 
Paolo Pani$^{3,1}$
}
\affiliation{${^1}$ CENTRA, Departamento de F\'{\i}sica, Instituto Superior T\'ecnico, Universidade de Lisboa, Avenida~Rovisco Pais 1, 1049 Lisboa, Portugal}
\affiliation{${^2}$ Perimeter Institute for Theoretical Physics, 31 Caroline Street North Waterloo, Ontario N2L 2Y5, Canada}
\affiliation{${^3}$ Dipartimento di Fisica, ``Sapienza'' Universit\`a di Roma \& Sezione INFN Roma1, Piazzale Aldo Moro 5, 00185, Roma, Italy}

\begin{abstract}
The existence of black holes and of spacetime singularities is a fundamental issue in science. Despite this, observations supporting their existence are scarce,
and their interpretation unclear. We overview how strong a case for black holes has been made in the last few decades, and how well observations adjust to this paradigm.
Unsurprisingly, we conclude that observational proof for black holes is impossible to come by. However, just like Popper's black swan, alternatives can be ruled out or confirmed to exist with a single observation. These observations are within reach. In the next few years and decades, we will enter the era of precision gravitational-wave physics
with more sensitive detectors. Just as accelerators require larger and larger energies to probe smaller and smaller scales, more 
sensitive gravitational-wave detectors will be probing regions closer and closer to the horizon, potentially reaching Planck scales and beyond. What may be there, lurking?
\end{abstract}


\maketitle

\noindent{\bf{\em I. Introduction.}}
The discovery of the electron and the known neutrality of matter led in 1904 to J. J. Thomson's ``plum-pudding'' atomic model. Data
from new scattering experiments was soon found to be in tension with this model, which was eventually superseeded by Rutherford's, featuring an atomic nucleus.
The point-like character of elementary particles opened up new questions. How to explain the apparent stability of the atom? How to handle the singular behavior of
the electric field close to the source? What is the structure of elementary particles? Some of these questions were elucidated with quantum mechanics and quantum field theory. Invariably, the path to the answer led to the understanding of hitherto unknown phenomena. 
The history of elementary particles is a timeline of the understanding of the electromagnetic (EM) interaction, and is pegged to its characteristic $1/r^2$ behavior (implying that other structure {\it has} to exist on small scales within any sound theory).
Arguably, the elementary particle of the gravitational interaction are black holes (BHs). Within General Relativity (GR), BHs are indivisible and are in fact the simplest macroscopic objects that one can conceive.

Can BHs hold the same surprises that the electron and the hydrogen atom did when they started to be experimentally probed?
Are there any parallels that can be useful guides? 
The BH interior is causally disconnected from the exterior by an event horizon. Unlike the classical description of atoms, the GR description of the BH exterior is self-consistent and free of pathologies. 
The ``inverse-square law problem'' --~the GR counterpart of which is the appearance of pathological curvature singularities~-- is swept to inside the horizon and therefore harmless for the external world. 
There are good indications that classical BHs are perturbatively stable against small fluctuations, and attempts to produce naked singularities, starting from generic initial conditions, have failed. BHs are not only curious mathematical solutions to Einstein's equations, but also their {\it formation} process is well understood.
In fact, there is nothing spectacular with the presence or formation of an event horizon. The equivalence principle dictates that an infalling observer crossing this region (which, by definition, is a \emph{global} concept)
feels nothing extraordinary: in the case of macroscopic BHs 
all of the local physics at the horizon is rather unremarkable.

Black holes have become so entrenched in scientific culture that, currently, an informal definition of a BH might well be ``any dark, compact object with mass above three solar masses.''
Why, then, should one question the existence of BHs?
There are a number of important reasons to do so. 
The BH exterior is pathology-free, but the interior is not. BHs harbour physical singularities or Cauchy horizons, which prevent the {\it prediction}
of the future of a regular, well-defined initial state. In fact, the geometry describing the interior of an astrophysical spinning BH is currently unknown. A possible resolution of this problem may require accounting for quantum effects. However, these lead to Hawking radiation which is at the root of the information loss puzzle.
The resolution of such problems could include changing the endstate of collapse~\cite{Mazur:2004fk,Mathur:2005zp,Mathur:2008nj} (perhaps as-yet-unknown physics can prevent the formation of horizons), 
or altering drastically the near-horizon region~\cite{Unruh:2017uaw}.
Such possibility is not too dissimilar from what happened with the atomic model after the advent of quantum electrodynamics.

Horizons are not only a rather generic prediction of GR, but their existence is in fact \emph{necessary} for the consistency of the theory at the classical level. This is the root of Penrose's (weak) Cosmic Censorship Conjecture, which remains one of the most urgent open problems in fundamental physics. In this sense, the statement that there is a horizon in any spacetime harboring a singularity in its interior is such a remarkable claim, that it requires similar remarkable evidence.
This is the question we will entertain here, {\it is there any observational evidence for the existence of BHs?}

Scientific paradigms have to be constantly questioned and subjected to experimental and observational scrutiny.  If the answer turns out to be that BHs do not exist, the consequences are so extreme and profound, that it is worth all the possible burden of actually testing it. Within the coming years we will finally be in the position of performing unprecedented tests on the nature of compact dark objects, the potential pay-off of which is enormous. As we argue, the question is not just whether the strong-field gravity region near compact objects is consistent with GR predictions, but rather to \emph{quantify} the limits of observations in testing the event horizon.

Known physics all but exclude BH alternatives. Nonetheless, the Standard Model of fundamental interactions
is not sufficient to describe the cosmos --~at least on the largest scales~-- and also leaves all of the fundamental questions regarding BHs open. It may be wise to keep an open mind.

\noindent{\bf{\em II. Photospheres and compact objects.}}
We focus on spherical symmetry for simplicity. Then, a vacuum BH is described by the 
Schwarzschild geometry, and has an event horizon at a location $r_g=2GM/c^2$, where $G$ is Newton's constant, $c$ the speed of light and $M$ the BH mass.
We will entertain the possibility that there exist {\it Exotic Compact Objects}~(ECOs), more massive than neutron stars but without horizons, with an effective surface at
\begin{equation}
r_0=r_g(1+\epsilon)\,,
\end{equation}
with $\epsilon \ll 1$. Although the above definition is coordinate-dependent, the proper distance between the surface and $r_g$ scales like $\epsilon^{1/2}$.
All the results discussed below show a dependence on $\log \epsilon$, making the distinction irrelevant. For Planckian corrections of an astrophysical BH, $\epsilon \sim 10^{-40}$. The motion of light or other massless particles around these objects shows two interesting features. Light emitted from the surface of ECOs is strongly lensed back towards it. In fact, all light rays fall back on the object on a time scale $\sim 5 r_g/c$, with the exception of those emitted almost radially within a solid angle
$\Delta \Omega_{\rm esc} \sim 27\pi\epsilon/4$. 
Any injected energy $\delta E\ll Mc^2$ is eventually radiated in a timescale longer than the Hubble time whenever 
\begin{equation}
 \epsilon\ll 10^{-16}\left(\frac{M}{10^6 M_\odot}\right)\,. \label{condequilibrium}
\end{equation}
Thus, very compact ECOs tend to be dark, just like BHs, even when surrounded by accretion disks.

Another truly relativistic feature of ECOs is that high frequency EM or gravitational waves (GWs) can orbit them in a circular motion. Newtonian theory does not allow for this, in GR it is only possible at a specific location, $r=\frac{3}{2} r_g$, which defines a surface called the photosphere. By its very own nature, the photosphere controls how BHs or ECOs look like when illuminated by accretion disks or stars, thus defining their so-called ``shadow''. Imaging these shadows for the supermassive dark source SgrA$^*$
at the center of our galaxy is the main goal of the Event Horizon Telescope~\cite{Loeb:2013lfa,Goddi:2016jrs}. The photosphere also has a bearing on the spacetime response to any type of waves, and therefore describes how high-frequency GWs linger close to the horizon. The circular motion of light rays is unstable: a small displacement grows exponentially on a timescale $\sim 3\sqrt{3}r_g/2c \approx 2.5r_g/c$~\cite{Ferrari:1984zz,Cardoso:2008bp}.
A geodesic description anticipates that light or GWs persist at the photosphere on these timescales.
Because the geodesic calculation is local, these conclusions hold irrespectively of the spacetime being vacuum all the way to the horizon or not. An ultracompact object features exactly the same geodesics and properties close to its photosphere, provided that on timescales $\approx 7.5r_g/c$ the null geodesics did not have time to bounce off the surface.
We are requiring three $e$-fold times for the instability to dissipate more than $99.7\%$ of the energy of the initial pulse. This amounts to requiring that
\be
\epsilon\lesssim 0.0165\,. \label{eps_crit}
\ee
Thus, the horizon plays no special role in the response of high frequency waves, nor could it: it takes an infinite (coordinate) time for a light ray to reach the horizon.
The above threshold on $\epsilon$ is a natural sifter between two classes of compact, dark objects. 
For objects characterized by $\epsilon \gtrsim 0.0165$, light or GWs can make the roundtrip from the photosphere to the object's surface and back, before dissipation of the photosphere modes occurs. For objects satisfying \eqref{eps_crit}, the waves trapped at the photosphere relax away by the time that the waves from the surface hit it back.

\begin{figure*}[ht]
\begin{center}
\includegraphics[width=1\textwidth]{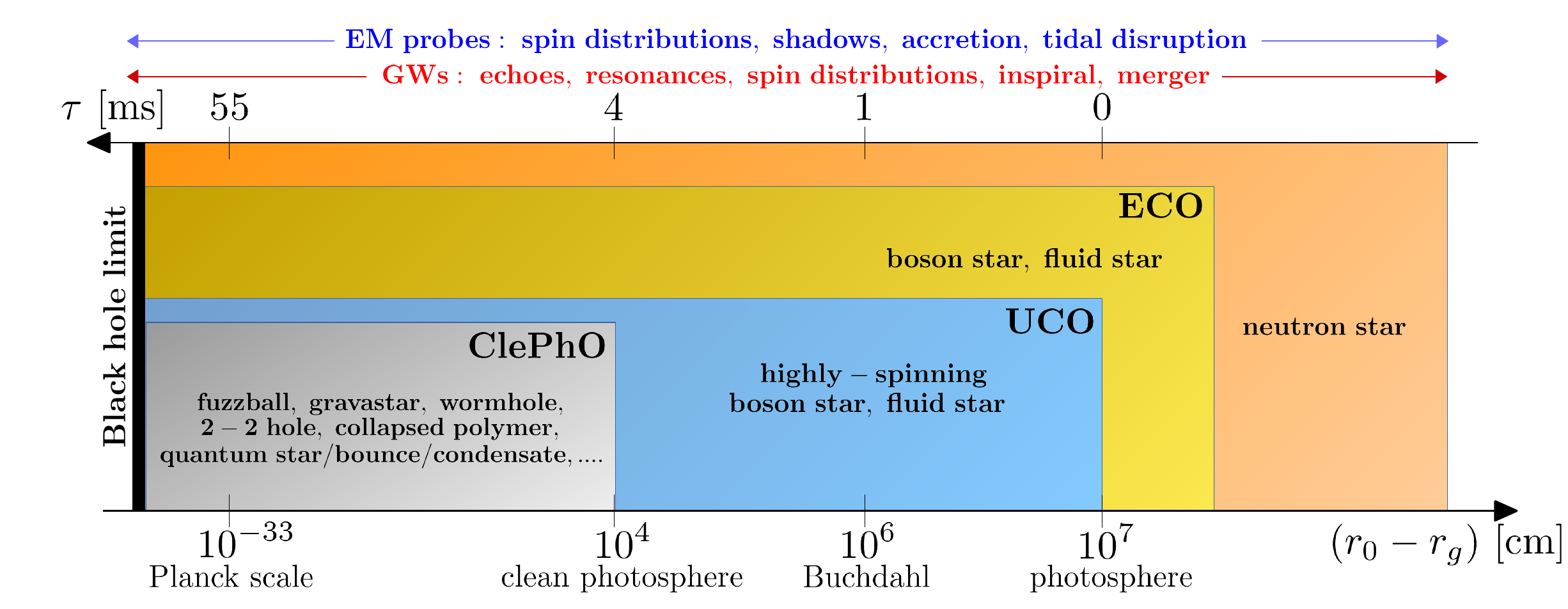}
\caption{\emph{Schematic classification of dark compact objects}. Their compactness is expressed as the difference between the object radius $r_0$ and the Schwarzschild radius $r_g$. Objects in the same category have similar dynamical properties on a timescale 
$\tau\sim \frac{r_g}{c}|\log\epsilon|$. The upper axis refers to the time, as measured by distant observers, that light from the photosphere takes to reach the surface $r_0$. Numbers refer to an object of $60M_\odot$ and scale linearly with it mass.
\label{fig:diagram}}
\end{center}
\end{figure*} 
We can thus use the properties of the photosphere to distinguish between different classes of models. 
Among the compact objects, some feature photospheres. These could be called \emph{ultracompact objects} (UCOs).
Those objects which satisfy condition \eqref{eps_crit} have a ``clean'' photosphere, and will be designated by \emph{ClePhOs}.
The early-time dynamics of ClePhOs is expected to be the same as that of BHs. At late times, ClePhOs should display unique signatures of their surface. The zoo of compact objects is summarized in Fig.~\ref{fig:diagram}. 

\noindent{\bf{\em III. Prompt ringdown and echoes.}}
\begin{figure}[ht]
\begin{center}
\includegraphics[width=0.49\textwidth]{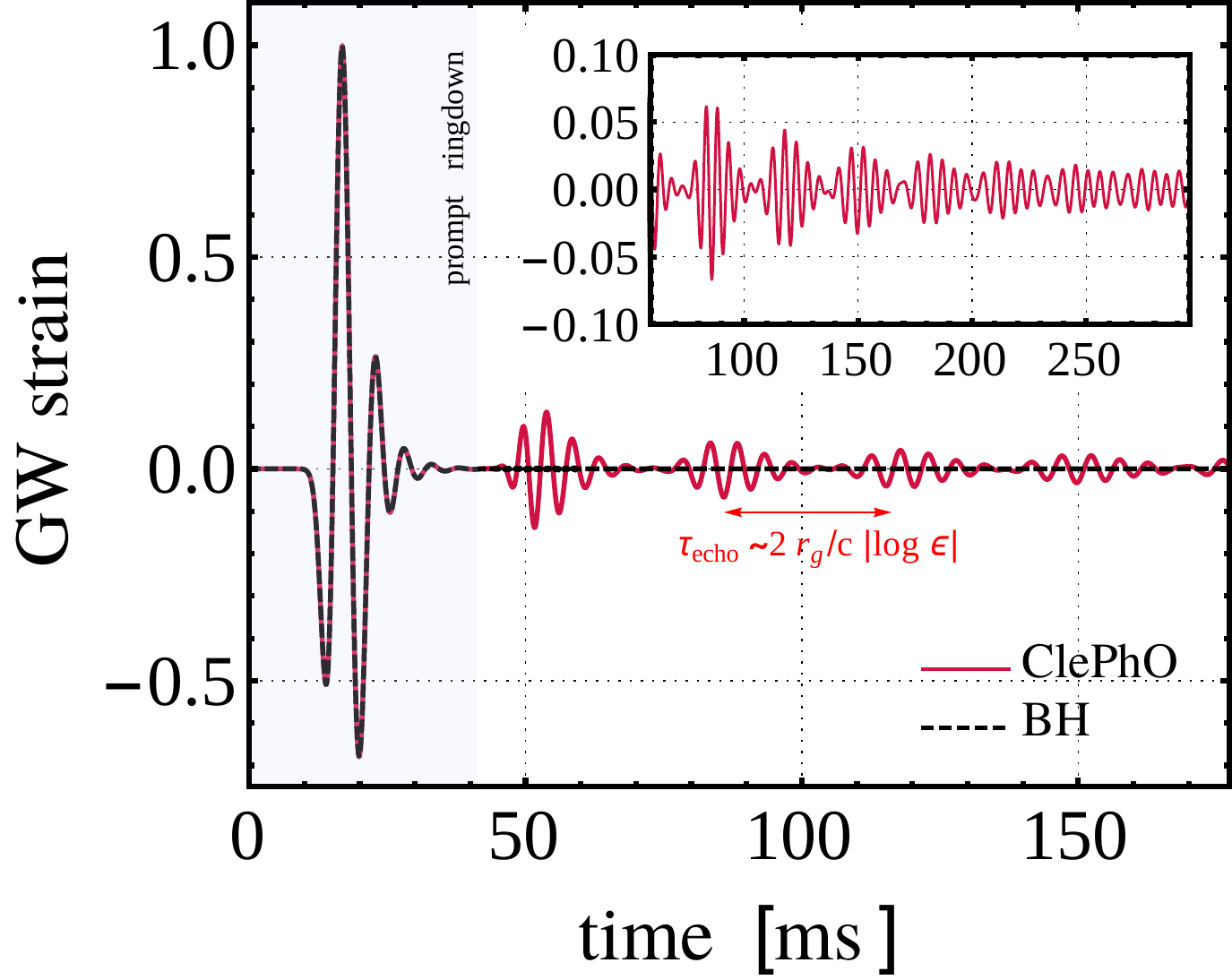}
\caption{\emph{Ringdown waveforms from black holes (black line) and ClePhOs (red line)}. We consider objects of $60M_\odot$. For ClePhOs, there is a reflective surface at $r_0=r_g(1+\epsilon)$, $\epsilon=10^{-11}$. The amplitude of the GW signal (proportional to the relative strain of the interferometer's arm induced by the GW) is normalized to its peak value. The initial data describes a quadrupolar Gaussian wavepacket of axial GWs. The inset shows a zoom-in version of the waveform at late times. Note that each subsequent echo has a smaller frequency content.
\label{fig:ringdown}}
\end{center}
\end{figure}
The above intuitive picture works reasonably well, and can be improved by doing a complete analysis of how GWs propagate around BHs or ECOs.
Take, for example, a BH and throw a pulse of low-frequency GWs at it. A fraction of those waves is reflected by the centrifugal barrier and scatters promptly off to infinity. The remaining, however, 
lingers longer at the photosphere before leaking away. The correct numbers for the frequency and damping time of quadrupolar waves are in rough agreement with the previous analysis and yield~\cite{Berti:2009kk}
\be
f=12.07\left(\frac{M_{\odot}}{M}\right) {\rm kHz}\,,\,\,\tau=55.37 \left(\frac{M}{M_{\odot}}\right) \mu{\rm s}\,.\label{bh_qnm}
\ee
These are the (quasi)normal modes of the system. The structure of GW signals at late times is therefore expected to be relatively simple. This is shown in Fig.~\ref{fig:ringdown}, 
which refers to the scattering of a Gaussian pulse off a BH. The pulse crosses the photosphere, and excites its modes. The ringdown signal, a fraction of which travels to outside observers, is to a very good level described by its lowest modes, Eq.~\eqref{bh_qnm}. 
The fraction of the GWs that leaks from the barrier {\it inwards} travels down to the horizon and that's the last one hears of it.

Contrast the previous description with the dynamical response of a ClePhO. The initial evolution of the photosphere modes still holds, by causality. Thus, up to timescales of the order $\sim \frac{r_g}{c}|\log\epsilon|$ (the roundtrip time of radiation between the photosphere and the surface) the signal is {\it identical} to that of BHs~\cite{Cardoso:2016rao,Cardoso:2016oxy}.
At later times, however, the pulse traveling inwards is bound to interact with the object. This pulse is semi-trapped between the object and the photosphere. Upon each interaction, a fraction exits to outside observers, giving rise to a series of {\it echoes} of ever-decreasing amplitude. Repeated reflections occur in a characteristic echo delay time~\cite{Cardoso:2016rao,Cardoso:2016oxy},
\begin{equation}
\tau_{\rm echo}\sim \frac{2r_g}{c} |\log\epsilon|\,. \label{tauecho}
\end{equation}
This logarithmic dependence is crucial to make echoes observable even with only Planckian corrections near the horizon, when $\epsilon\sim10^{-40}$.
Although, at very late times, the fundamental modes of a ClePhO have low frequencies, the main burst is typically generated at the photosphere and has therefore a frequency content of the same order as the BH modes~\eqref{bh_qnm}.
The initial signal is of high frequency and a substantial component is able to cross the potential barrier. Thus, observers see a series of echoes whose amplitude is getting smaller and whose frequency content is also going down (see Fig.~\ref{fig:ringdown}).

\noindent{\bf{\em IV. Beyond vacuum black holes.}}
To entertain the possibility that dark, massive, compact objects are not BHs, requires
one to discuss some outstanding issues. One can take two different stands on this topic:
(i) a pragmatic approach of testing the spacetime close to these objects, irrespective
of their nature, by devising model-independent observations that yield unambiguous answers;
(ii) a less ambitious and more theoretically-driven approach, which starts by constructing horizonless objects, within some framework. It proceeds to study their formation mechanisms and stability properties; then discard solutions which either do not form or are unstable on relatively short timescales; finally, understand the observational imprints of the remaining objects, and how they differ from BHs'.

In practice, when dealing with outstanding problems where our ignorance is extreme, both approaches should be used simultaneously.
Indeed, using concrete models can sometimes be a useful guide to learn about broad, model-independent signatures.
While the statement ``BHs exist in our Universe'' is \emph{fundamentally unfalsifiable},
alternatives can be ruled out or confirmed to exist with a single observation, just like Popper's black swans.

A nonexhaustive summary of possible objects which could mimic BHs is shown in Fig.~\ref{fig:diagram}. Stars made of constant density fluids are perhaps the first known example of compact configurations.
In GR, isotropic static spheres made of ordinary fluid satisfy the Buchdahl limit on their compactness, $r_g/r_0<8/9$~\cite{Buchdahl:1959zz}, and can thus never be a ClePhO.

Compact solutions can be built with fundamental, massive bosonic fields~\cite{Kaup:1968zz,Ruffini:1969qy,Seidel:1991zh,Brito:2015yga,Liebling:2012fv}. The quest for self-gravitating  structures started in attempts to understand if ``solitons'' could arise out of the non-linearities of Einstein field equations. Such configurations are broadly referred to as {\it boson stars}, and have attracted considerable interest since scalars can explain, for instance, the dark matter puzzle~\cite{Hui:2016ltb}. When fastly spinning, such stars can be extremely compact. There seem to be no studies on the classification of such configurations (there are solutions known to display photospheres, but it is unknown whether they can be as compact as ClePhOs).

The other objects in the list require either unknown matter or large quantum effects. These include {\it wormholes}, an example
of which was discussed carefully in the context of ``BH foils''~\cite{Damour:2007ap}; {\it fuzzballs}~\cite{Mathur:2005zp,Mathur:2008nj}, were introduced as a microstate description of BHs in string theory. In this setup the individual microstates
are horizonless and the horizon arises as a coarse-grained description of the microstate geometries. 
Phenomena such as Hawking radiation can be recovered, in some instances,
from classical instabilities~\cite{Chowdhury:2007jx}.
``Gravitational-vacuum stars'' or {\it gravastars} are ultracompact configurations supported by a negative pressure, which might arise 
as an hydrodynamical description of one-loop quantum effects in curved spacetime~\cite{Mazur:2004fk,Mottola:2006ew}.

For most objects which are inspired by quantum-gravity corrections, the changes in the geometry occur
deep down in the strong field regime. Some of these models --~including also {\it black stars}~\cite{Barcelo:2009tpa,Barcelo:2007yk}, {\it superspinars}~\cite{Gimon:2007ur} or {\it collapsed polymers}~\cite{Brustein:2016msz,Brustein:2017kcj}~-- predict that large quantum backreaction should affect the horizon geometry even for macroscopic objects. In these models, the parameter $\epsilon$ is naturally of the order $\sim 10^{-39}-10^{-46}$ for masses in the range $10-10^8 M_\odot$, or even smaller in some models~\cite{Holdom:2016nek}.
In all these cases, quantum-gravity or microscopic corrections at the horizon scale select ClePhOs as well-motivated alternatives to BHs.

Although supported by sound arguments, the vast majority of the alternatives to BHs are, at best, incompletely described. Precise calculations (and often even a rigorous framework) incorporating the necessary physics are still missing. One notable exception to our ignorance are boson stars. These configurations are known to arise, generically, out of the gravitational
collapse of massive scalars. Their interaction and mergers can be studied by evolving the Einstein-Klein-Gordon system, and there is evidence that accretion of less massive boson stars
makes them grow and cluster around the configuration of maximum mass. In fact, boson stars have efficient {\it gravitational cooling}
mechanisms that allow them to avoid collapse to BHs and remain very compact after interactions~\cite{Seidel:1991zh,Seidel:1993zk,Brito:2015yfh}.

It is common lore that ClePhOs are so compact that their merger must result in a BH. However, if large quantum effects do occur, they would probably act on short timescales
to prevent apparent horizon formation. In some models, Planck-scale dynamics naturally leads to abrupt changes close to the would-be horizon, without fine tuning.

\noindent{\bf{\em V. On the stability of ultracompact objects.}}
Appealing solutions are only realistic if they form and remain as long-term stable solutions of the theory.
There are good reasons to believe that some --~if not all~--
horizonless compact solutions are unstable. 
We would like to highlight two general results. Linearized gravitational fluctuations of any nonspinning UCO are extremely long-lived and decay no faster than logarithmically~\cite{Keir:2014oka,Cardoso:2014sna}. The long damping time of these modes has led to the conjecture that any UCO is nonlinearly unstable and may evolve through a Dyson-Chandrasekhar-Fermi type of mechanism~\cite{Keir:2014oka,Cardoso:2014sna}. The endstate is unknown, and most likely depends on the equation of the state of the particular UCO: some objects may fragment and evolve past the UCO region into less compact configurations, via mass ejection, whereas other UCOs may be forced into gravitational collapse to BHs.

The above mechanism is supposed to be active for any spherically symmetric UCO, and also on spinning solutions. However, it is nonlinear in nature. On the other hand, UCOs (and especially ClePhOs) can develop negative-energy regions once spinning. In such a case, there is a well-known {\it linear} instability, affecting any horizonless geometry with an ergoregion, the final state of which is likely a slowly-spinning ClePhO~\cite{Friedman:1978wla,Moschidis:2016zjy,Cardoso:2007az,Maggio:2017ivp}.

Unfortunately, the effect of viscosity is practically unknown~\cite{Cardoso:2014sna},
and so are the timescales involved in putative dissipation mechanisms that might quench this instability. On the other hand, even unstable solutions are relevant if the timescale is larger than any other dynamical scale in the problem. 
The nonlinear mechanism presumably acts on very long timescales only; a model problem predicts an exponential dependence on the size of the initial perturbation~\cite{FritzJohn}.
For ClePhOs, the ergoregion-instability timescale is parametrically longer than the dynamical timescale of the object~\cite{Friedman:1978wla,Cardoso:2007az,Maggio:2017ivp}.
Given such long timescales, it is likely that the instability can be efficiently quenched by some dissipation mechanism of nongravitational nature, although this effect would be model-dependent~\cite{Maggio:2017ivp}.

Finally, there are (surprising!) indications that classical instabilities of UCOs are merely the equivalent of Hawking radiation for these geometries, and that therefore there might be a smooth transition in the emission properties when approaching the BH limit~\cite{Chowdhury:2007jx,Damour:2007ap}.

\noindent{\bf{\em VI. The gravitational-wave era.}}
The historical detection of GWs by aLIGO~\cite{Abbott:2016blz} opens up the exciting possibility of testing gravity in extreme regimes with unprecedented accuracy~\cite{TheLIGOScientific:2016src,Yunes:2013dva,Barausse:2014tra,Berti:2015itd,Yunes:2016jcc,Maselli:2017cmm}. GWs are generated by coherent motion of massive sources, and
are therefore subjected to less modeling uncertainties (they depend on far fewer parameters) relative to EM probes. The most luminous GWs come from very dense sources, but they also interact very feebly with matter, thus providing the cleanest picture of the cosmos, complementary to that given by telescopes and particle detectors.

Compact binaries are the preferred sources for GW detectors~\cite{Abbott:2016blz,Abbott:2016nmj}. The GW signal from compact binaries is naturally divided
in three stages, corresponding to the different cycles in the GW-driven evolution~\cite{Buonanno:2006ui,Berti:2007fi,Sperhake:2011xk}:
the inspiral stage, corresponding to large separations and described by post-Newtonian theory; the merger phase when the two objects coalesce and which can only be described accurately through numerical simulations; and finally, the ringdown phase when the
merger end-product relaxes to a stationary, equilibrium solution of the field equations~\cite{Sperhake:2011xk,Berti:2009kk,Blanchet:2013haa}.

All three stages provide independent, unique tests of gravity and of compact GW sources.
The GW signal in the events reported so far is consistent with them being generated by BH binaries,
and with the endstate being a BH. To which level, and how, are alternatives consistent with current and future observations?
This question may be divided into two different schemes.

\noindent{\bf{\em VIa. Smoking guns: echoes and resonances.}}
For binaries composed of ClePhOs, the GWs generated during inspiral and merger is expected to be very similar to
those from a corresponding BH binary with the same mass and spin. However, clear distinctive features appear due to the absence of a horizon. The most prominent signature is the appearance of late-time echoes in the waveforms~\cite{Cardoso:2016rao,Cardoso:2016oxy,Price:2017cjr,Nakano:2017fvh}. After the merger, ClePhOs give rise to two different ``ringdown'' signals: the first stage is dominated by the photosphere modes, and is indistinguishable from a BH ringdown. After a time $\sim\tau_{\rm echo}$, the waves
trapped in the ``photosphere+ClePhO'' cavity start leaking out as echoes of the main burst as in Fig.~\ref{fig:ringdown}. This is a smoking-gun for new physics, potentially reaching microscopic or even Planckian corrections at the horizon scale~\cite{Cardoso:2016rao,Cardoso:2016oxy}.

Strategies to dig GW signals containing ``echoes'' out of noise are not fully under control, first efforts are underway [A. Nielsen and C. Van den Broeck; private communications]. 
The ability to detect such signals depends on how much energy is converted from the main burst into echoes (i.e., on the relative amplitude between the first echo and the prompt ringdown signal in Fig.~\ref{fig:ringdown}).
Define the ratio of energies to be $\gamma_{\rm echo}$. Then, the signal-to-noise ratio $\rho$ necessary for echoes to be detectable {\it separately} from the main burst, for a detection threshold of $\rho=8$, is
$\rho_{\rm prompt\, ringdown}\gtrsim \frac{80}{\sqrt{\gamma_{\rm echo}(\%)}}$.
In such case the first echo is detectable with a simple ringdown template (an exponentially damped sinusoid~\cite{Abbott:2009km}).
Space-based detectors will see prompt ringdown events with very large $\rho$~\cite{Berti:2016lat}. For $\gamma_{\rm echo}=20\%$, we estimate that the planned space mission LISA~\cite{Audley:2017drz} will see at least one event per year, even for the most pessimistic population synthesis models used to estimate the rates~\cite{Berti:2016lat}. The proposed Einstein telescope~\cite{Punturo:2010zz} or Voyager-like~\cite{Voyager} third-generation Earth-based detectors will also be able to distinguish ClePhOs from BHs with such simple-minded searches.
The event rates for LIGO are smaller, and more sophisticated searches need to be implemented. Preliminary analysis of GW data using the entire echoing sequence was reported recently~\cite{Abedi:2016hgu,Ashton:2016xff,Abedi:2017isz}. 
More detailed characterization of the echo waveform (e.g., using Green's function techniques~\cite{Mark:2017dnq} and accounting for spin effects~\cite{Nakano:2017fvh}) is necessary to reduce the systematics in data analysis.

Matter modes may also be excited. Due to redshift effects, these presumably play a subdominant role in the GW signal. If ECOs arise in effective theories with extra gravitational degrees of freedom~\cite{Mottola:2016mpl}, extra polarization modes will be present.
The possible detection of these modes with a global network of interferometers could be another indication of new physics at the horizon scale.

In addition, resonant mode excitation during inspiral may occur. The echoes at late times are just vibrations of the ClePhO. These vibrations have relatively low frequency and can, in principle, also be excited during the inspiral process itself, leading to resonances in the motion as a further clear-cut signal of new physics~\cite{Pani:2010em,Macedo:2013qea,Macedo:2013jja}. 

Finally, the ergoregion instability depletes ClePhOs of their angular momentum: GW (or EM) observations indicating statistical prevalence of slowly-spinning objects, across the entire mass range, could be an indication for the absence of an horizon. Assuming that the exterior geometry of ClePhOs is well described by the Kerr metric, an object with dimensionless spin $\chi\gtrsim0.2$ and $\epsilon\sim 10^{-43}$ would have an instability timescale $\tau<3\times10^5 [M/(60M_\odot)]\,{\rm s}$.

\noindent{\bf{\em VIb. Precision physics.}}
The GW astronomy era will also gradually open the door to precision physics, for which smoking signs may not be necessary to test new physics. If the product of the merger is an ECO but not a ClePhO, it will simply vibrate differently from a BH; precise measurements of the ringing frequency and damping time allow one to test whether or not the object is a BH~\cite{Berti:2005ys,Berti:2006qt}.
Such tests are in principle feasible for wide classes of objects, including boson stars~\cite{Berti:2006qt,Macedo:2013qea,Macedo:2013jja}, gravastars~\cite{Chirenti:2016hzd}, wormholes~\cite{Konoplya:2016hmd,Nandi:2016uzg}, or other quantum-corrected objects~\cite{Barcelo:2017lnx,Brustein:2017koc}.

The merger phase can also provide information about the nature of the coalescing objects. ECOs which are not sufficiently compact will likely display a merger phase resembling that of a neutron-star rather than a BH merger~\cite{Cardoso:2016oxy,Bezares:2017mzk}. The situation for ClePhOs is unclear since no simulations of a full coalescence are available. It is also possible that, at variance with the BH case, the merger of two ClePhOs can be followed by a burst of EM radiation associated with the presence of high-density matter during the collision. Although such emission might be strongly redshifted, searches for EM counterparts of candidate BH mergers can provide another distinctive signature of new physics at the horizon scale.

The absence of a horizon affects the way in which the inspiral stage proceeds. In particular, three different features may play a role, all of which can be used to test the ``BHness'' of the objects: \emph{(i)} There is no tidal heating for ECOs. Horizons absorb incoming high-frequency GWs, and serve as sinks or amplifiers for low-frequency radiation able to tunnel in.
UCOs and ClePhOs, on the other hand, are not expected to absorb any significant amount of GWs (although the trapping of radiation in ClePhOs can efficiently mimic the effect of a horizon  when $\epsilon$ is sufficiently small~\cite{Cardoso:2017cfl}). Thus, a ``null-hypothesis'' test consists on using the phase of GWs to measure absorption or amplification at the surface of the objects. 
Since supermassive binaries produce louder signals, LISA-type GW detectors may place stringent tests on this property, potentially reaching Planck scales near the horizon and beyond~\cite{Cardoso:2017cfl,Maselli:2017cmm}. 
\emph{(ii)} ECOs are deformable. In a binary, the gravitational pull of one object deforms its companion, inducing a quadrupole moment proportional to the tidal field. The tidal deformability is encoded in the Love numbers, and the consequent modification of the dynamics can be directly translated into the GW phase evolution~\cite{Flanagan:2007ix,Maselli:2017cmm}. 
A remarkable result in classical GR states that the Love numbers of BHs are zero~\cite{Binnington:2009bb,Damour:2009vw,Poisson:2014gka,Pani:2015hfa}, allowing again for null tests. Future LIGO observations have the potential to distinguish BHs from boson stars~\cite{Wade:2013hoa,Cardoso:2017cfl,Sennett:2017etc}, or even generic ClePhOs, for which the Love numbers scale logarithmically with $\epsilon$~\cite{Cardoso:2017cfl}.
\emph{(iii)} Different multipolar structure. Spinning ECOs possess multipole moments that differ from those of BHs. The impact of the multipolar structure on the GW phase allows one to estimate and constrain possible deviations from the BH paradigm~\cite{Krishnendu:2017shb}.

Thomson's atomic model was carefully constructed, and tested theoretically for inconsistencies.
Rutherford's incursion into scattering of $\alpha$ particles was not meant to disprove the model, it was just aimed at testing its accuracy. According to Marsden, ``...it was one of those 'hunches' that perhaps some effect might be observed, and that in any case that neighbouring territory of this Tom Tiddler's ground might be explored by reconnaissance. Rutherford was ever ready to meet the unexpected and exploit it, where favourable, but he also knew when to stop on such excursions''~\cite{Birks}.

At one hand's reach, we have the possibility to dig deeper and deeper into the strong-field region of compact objects, {\it for free}. 
As the sensitivity of EM and GW detectors increases, so does our ability to probe 
regions of ever increasing redshift, closer and closer to the horizon. Perhaps the strong-field region of gravity holds the same surprises that the strong-field EM region did?

Correspondence and requests for materials should be addressed to \url{vitor.cardoso@ist.utl.pt}.

\noindent{\bf{\em Acknowledgments.}}
%
V.C. acknowledges financial support provided under the European Union's H2020 ERC Consolidator Grant ``Matter and strong-field gravity: New frontiers in Einstein's theory'' grant agreement no. MaGRaTh--646597.
Research at Perimeter Institute is supported by the Government of Canada through Industry Canada and by the Province of Ontario through the Ministry of Economic Development $\&$
Innovation.
This article is based upon work from COST Action CA16104 ``GWverse'', and MP1304 ``NewCompstar'' supported by COST (European Cooperation in Science and Technology).
This work was partially supported by FCT-Portugal through the project IF/00293/2013, by the H2020-MSCA-RISE-2015 Grant No. StronGrHEP-690904.

VC and PP contributed equally to the writing and calculations in this work.


\end{document}